\documentclass[aip,preprint,graphicx]{revtex4-1}

\usepackage[table,xcdraw]{xcolor}
\usepackage{amsmath}
\usepackage{graphicx}
\usepackage{float}
\usepackage{booktabs}
\usepackage{multirow}

\newcommand{\hydra}{\texttt{Hydra}}
\newcommand{\cretin}{\texttt{Cretin}}

\draft 

\begin{document}


\title{Transfer Learning as a Method to Reproduce High-Fidelity NLTE Opacities in Simulations} 



\author{Michael D. Vander Wal}
\email[]{mvander5@nd.edu}
\affiliation{University of Notre Dame}

\author{Ryan G. McClarren}
\email[]{rmcclarr@nd.edu}
\affiliation{University of Notre Dame}

\author{Kelli D. Humbird}
\email[]{humbird1@llnl.gov}
\affiliation{Lawrence Livermore National Laboratory}


\date{\today}

\begin{abstract}
Simulations of high-energy density physics often need non-local thermodynamic equilibrium (NLTE) opacity data. This data, however, is expensive to produce at relatively low-fidelity. It is even more so at high-fidelity such that the opacity calculations can contribute ninety-five percent of the total computation time. This proportion can even reach large proportions. Neural networks can be used to replace the standard calculations of low-fidelity data, and the neural networks can be trained to reproduce artificial, high-fidelity opacity spectra. In this work, it is demonstrated that a novel neural network architecture trained to reproduce high-fidelity krypton spectra through transfer learning can be used in simulations. Further, it is demonstrated that this can be done while achieving a relative percent error of the peak radiative temperature of the hohlraum of approximately 1\% to 4\% while achieving a 19.4x speed up.
\end{abstract}


\maketitle 

\section{Introduction}
Inertial confinement fusion (ICF) is one of the prospective methods for achieving commercial fusion power generation. While ICF research is on the cusp of achieving break even net energy output for the amount of energy provided by the laser drive, there is much work to still be done \cite{callahan2021}. One challenge associated with ICF energy production is that the experiments, or ``shots'', are expensive and limited in number. Simulations thus play an important role in selecting experimental parameters to maximize information gained from each shot. 

The simulations of ICF experiments have their own high costs, often taking several days on several nodes of the most powerful supercomputers in the world. For many integrated hohlraum simulations, the calculation of just the spectral absorptivity and emissivity, or opacity, can easily consume as much as ninety percent of the total computation time\cite{kluth}. This is, of course, dependent on the fidelity of the spectra being produced where the high-fidelity spectra compose 96\% of the entire computation time in this work. The higher the level of fidelity of the spectra being computed, the greater the proportion of the simulation time that is dedicated to the calculating the opacities.

While the higher cost of increased fidelity computations is to be expected, it does mean that there could be physics missing when running models at low fidelity. There can be significant changes in the spectra from low-fidelity to high-fidelity such that a band of frequencies, absorbed or emitted, may change in width, height or depth, and position. This is made rather apparent in previous work on the topic \cite{vanderwal2}. A simple example of an effect changes in model fidelity could have is the change in penetration depth, or the likelihood of escape or absorption of a photon. If that particular energy photon carries the majority of the energy emitted and the high-fidelity calculation results in an increase in that energy band, it is likely that less energy leakage will occur.

Neural networks have thus been suggested as an alternative to the normal usage of an atomic physics code such as \cretin{}. Previous work has demonstrated that, for a 1D simulation with a krypton hohlraum, using neural networks to perform the non-local thermodynamic equilibrium (NLTE) opacity calculations for krypton can net a 10x speedup in computation time with sub-one percent error for the radiation temperature during simulations \cite{kluth}. Furthermore, neural networks trained on high fidelity atomic physics data could enable not only faster, but more accurate calculations to be used in integrated ICF simulations. In this work, a technique called ``transfer learning'' is used to create high fidelity neural networks to predict NLTE opacities for use in integrated ICF simulations. Our results indicate the adoption of transfer learning allows high-fidelity NLTE physics to be included in routine ICF simulations at a computational cost comparable to low-fidelity calculations. 

\subsection{Prior Work and Contributions}
This work is an extension of previous work to replace low-fidelity krypton spectra in simulations run in the radiation hydrodynamics code \hydra\cite{kluth}. There has also been work on transfer learning to make neural networks that reproduce high-fidelity spectra from stand-in radiative fields and a large range of input values that include realistic and non-realistic combinations of inputs \cite{vanderwal2}. Other work related to the reproduction of spectra was the demonstration of a multi-element model and how the choice of data scaling is important and necessary to achieve accurate results \cite{vanderwal1}. 

Elsewhere in the fields of nuclear, atomic, and plasma physics have been a number of applications of machine learning. One of the more recent and substantial uses of machine learning in these fields has been the demonstration, at least by way of simulation, of the ability of neural networks to control magnetically confined plasmas in tokamak reactors \cite{degrave2022}. Neural networks are heavily used to improve the understanding of simulation discrepancies with experimental data and to create more accurate models to guide experiemental design\cite{humbird2020,humbird2021,kustowski2019}. In plasma physics, physics-informed neural networks have been used to evaluate plasma conditions around space craft as well as more general plasma behavior in space \cite{bard2021,zhang2021}.

Outside of the field of NLTE opacities or anything related to nuclear, atomic, or plasma physics there has been other work for the reproduction of spectra or the prediction of spectra for molecular compounds \cite{ghosh2019}. A more common application has been in the use of making calibration spectra for optical measurements \cite{lan2021neural,chatzidakis2019,cui2018,chen2019}. Not focusing on spectra prediction itself but rather the use of spectra, there has been work on the utilization of spectra to predict the properties of oil spills, the composition of plants, or properties of biological tissue \cite{liu2019,ni2019,lan2021neural}.

Finally, there is a plethora of papers addressing transfer learning. Transfer learning at the most general level is the process of taking a neural network trained on one set of data and then retraining it on similar, yet different data, of different levels of fidelity\cite{vanderwal2,aydin2019}. The idea behind transfer learning is that neural networks possess prior knowledge about a similar task, so the network can likely learn the new information with fewer training samples \cite{zhuang2021,meng2020,aydin2019,luo2018,zhu2019,zhuang2015,deng2013,deng2014,eusebio2018,imai2020}. Similarly, one can take part of an already trained network and attach a new network, or set of layers, to the pre-trained partial network. This particular process is quite common for networks working with images where a convolutional autoencoder is trained and then only convolutional filters in the encoder are used. The final product of these networks are used for classification \cite{luo2018,zhu2019,zhuang2015,deng2013,deng2014,eusebio2018}. Transfer learning is used for regression as well, and indeed, incorporating experimental data into a already trained network is a form of transfer learning \cite{meng2020,aydin2019,kustowski2019,goswami2020,humbird2021,humbird2020,chakraborty2021}. Some of the methods used can  become rather elaborate such as adding additional networks for linear and non-linear correlations for multi-fidelity transfer learning and requiring both levels of fidelity be utilized concurrently during training \cite{meng2020}.

The work provided here is an immediate follow up on the demonstration that high-fidelity spectra can be transfer learned in an effective fashion. In this work, we demonstrate the viability of using transfer learned models inside integrated ICF hohlraum simulations by reproducing the maximum radiation temperature observed within the hohlraum with mean relative error in the realms of 1\% with 94.8\% reduction in computation time for high-fidelity spectra. Previous work used  analytic radiative fields \cite{vanderwal2}; this work instead uses realistic radiative fields from Hydra simulations of ICF experiments that must be compressed with an autoencoder first before predictions can be made. This is similar to the work previously performed by Kluth et al., except we increase the resolution of the models and leverage transfer learning to improve the fidelity of the opacity calculations \cite{kluth}. Further differentiating it from the previous work, while not fully explored here, is the implementation of a novel neural network architecture and improved loss function for effective training.

\section{Methods}
\subsection{Data}
This work utilizes two separate datasets: one high-fidelity and one low-fidelity dataset. The low-fidelity model utilizes 1,849 levels, and the high-fidelity model utilizes 25,903 levels in atomic state. The models have sample rates of 35.3 samples/min and 3.2 samples/min respectively. Ultimately, this work uses 26,000 low-fidelity samples and 1,000 high-fidelity samples. In the following sections we outline to process of generating the data, and then discuss the training of the neural networks.


\subsubsection{Step One - Production of Radiative Field Data}
 An initial set of low-fidelity \hydra{} simulations are generated where \cretin{} provides the NLTE opacities, and Hydra produces radiative fields that are extracted for training. In addition to the input radiative fields, the  input temperature and density and the output absorptivity and emissivity spectra are also extracted for neural network training. In this work, 16,000 samples are obtained from the Hydra simulations.

\subsubsection{Step Two - Production of Low-Fidelity Absorptivity and Emissivity Data}
Second, \cretin{} produces an additional 10,000 samples of low-fidelity absorptivity and emissivity spectra of krypton using the radiative fields and randomly sampled density and electron temperatures as inputs. The spectra produced consist of 400 energy bins ranging from 0 eV to 40,000 eV. The density is sampled using a quasi-latin-hypercube sampling\cite{mcclarren2018uncertainty} weighted by the distribution of inputs obtained from the simulation data. The radiative fields are randomly sampled from a list of radiative fields that are binned together with the input values based on the integrated intensity of the radiative fields.

To provide some more clarity, imagine a 3D histogram that is binned with the electron temperature, the $\log_{10}$ transformation of the density, and the integrated intensity of the radiative fields in the three sampled dimensions and the counts of those bins in the fourth dimension. The bins with temperatures $>300$eV are more heavily weighted to increase the number of samples above 300 eV. The resultant weighting provides a roughly 50-50 split of temperature values above and below 300 eV as opposed to the 20-80 split that is actually present.

The  histogram is normalized and ordered in ascending order of the normalized counts with all bins containing a value of zero being excluded. A bin is then randomly sampled using latin-hypercube sampling. The density and temperature are then randomly sampled from random uniform distribution bounded by the edges of the bin. The radiative field is then sampled from a list of radiative fields that possess an integrated intensity that falls between the edges of the sampled bin. The binning is rather coarse to allow for the sampling of input values beyond the edge cases found in the simulation data. This is done to help improve the generalizability of the neural network beyond the simulations used to produce the radiative fields and other input values. Finally, the reason for not sampling from a random uniform distribution between two extrema is because that results in many input combinations that would not reasonably occur during an ICF simulation.

\subsubsection{Step Three - Production of High-Fidelity Absorptivity and Emissivity Data}
Third, \cretin{} further produces higher fidelity NLTE absorptivity and emissivity spectra utilizing the same process and distribution as in step two. The production of the higher fidelity spectra utilizes the same radiative fields from the low-fidelity \hydra{} simulations, since the premise of this study is that the high-fidelity simulations are too expensive to run hundreds of times. It should be noted that none of the absorptivity and emissivity samples directly from the simulations are used in the high-fidelity dataset because they are from low-fidelity models. Instead, the entire high-fidelity dataset is sampled using the prescribed sampling method.


\subsection{Process}
\subsubsection{Training}
\begin{figure}[H]
    \centering
    \includegraphics[width=\linewidth]{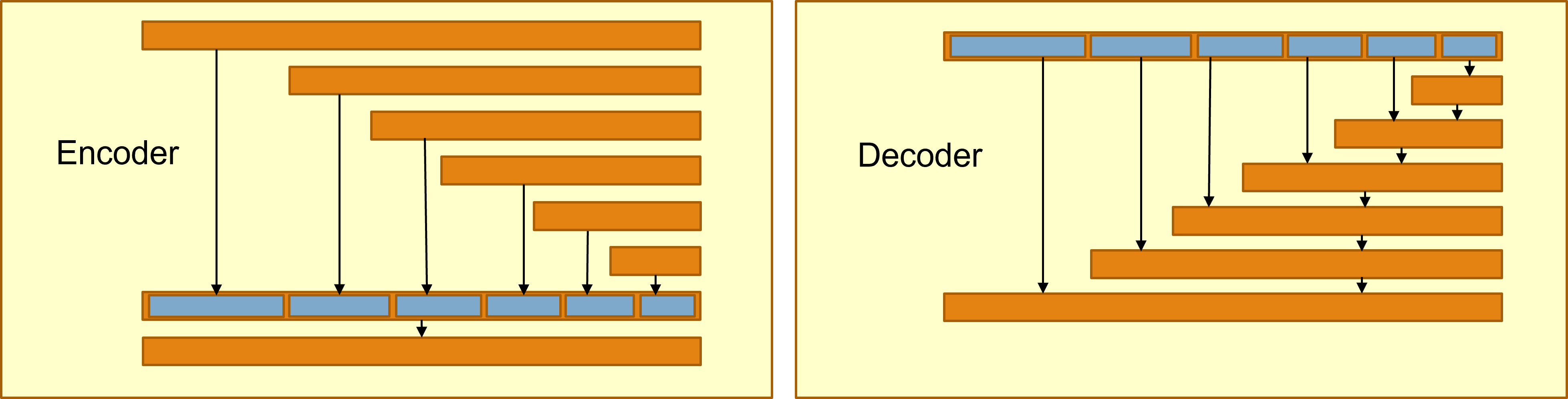}
    \caption{This is the general architecture of the encoder and decoder of the autoencoder. The latent space is a fully-connected layer that mixes the prior layer which contains sets of nodes that are directly connected to their own specific layers including the input and output.}
    \label{fig: multiscale ae}
\end{figure}

While this work is a direct continuation of the work in \cite{kluth,vanderwal2} it does not use the same  method of building a prediction model. Instead of creating separate autoencoders for the radiative fields and absorptivity and emissivity spectra that are purely sequential, fully-connected or convolutional neural networks, it utilizes a single, non-sequential, full-connected autoencoder for all radiative fields, absorptivity spectra, and emissivity spectra. The autoencoder is constructed as shown in Figure \ref{fig: multiscale ae} such that each layer of the encoder passes information to both the next layer in the encoder as well the layer just prior to the latent space. The decoder does the exact opposite.

Further, this work does not strictly use separate DJINN \cite{djinn} models for the absorptivity and emissivity neural network models as is done with the models in \cite{kluth,vanderwal1,vanderwal2} where a single DJINN model takes the radiative field latent space, density, temperature, atomic number as inputs to predict the latent space absorptivity or emissivity latent space. Instead, as seen in Figure \ref{fig: full arch}, two separate DJINN models are trained to predict the latent space of the absorptivity and emissivity spectra as encoded by the single, multiscale autoencoder by using only the material properties: density, temperature, and atomic number but not the radiative field which is treated as an environmental property. The density, temperature, and atomic number are minmax scaled. The two DJINN models are trained for 100 epochs and then combined in parallel into a single model.

\begin{figure}
    \centering
    \includegraphics[width=.75\textwidth]{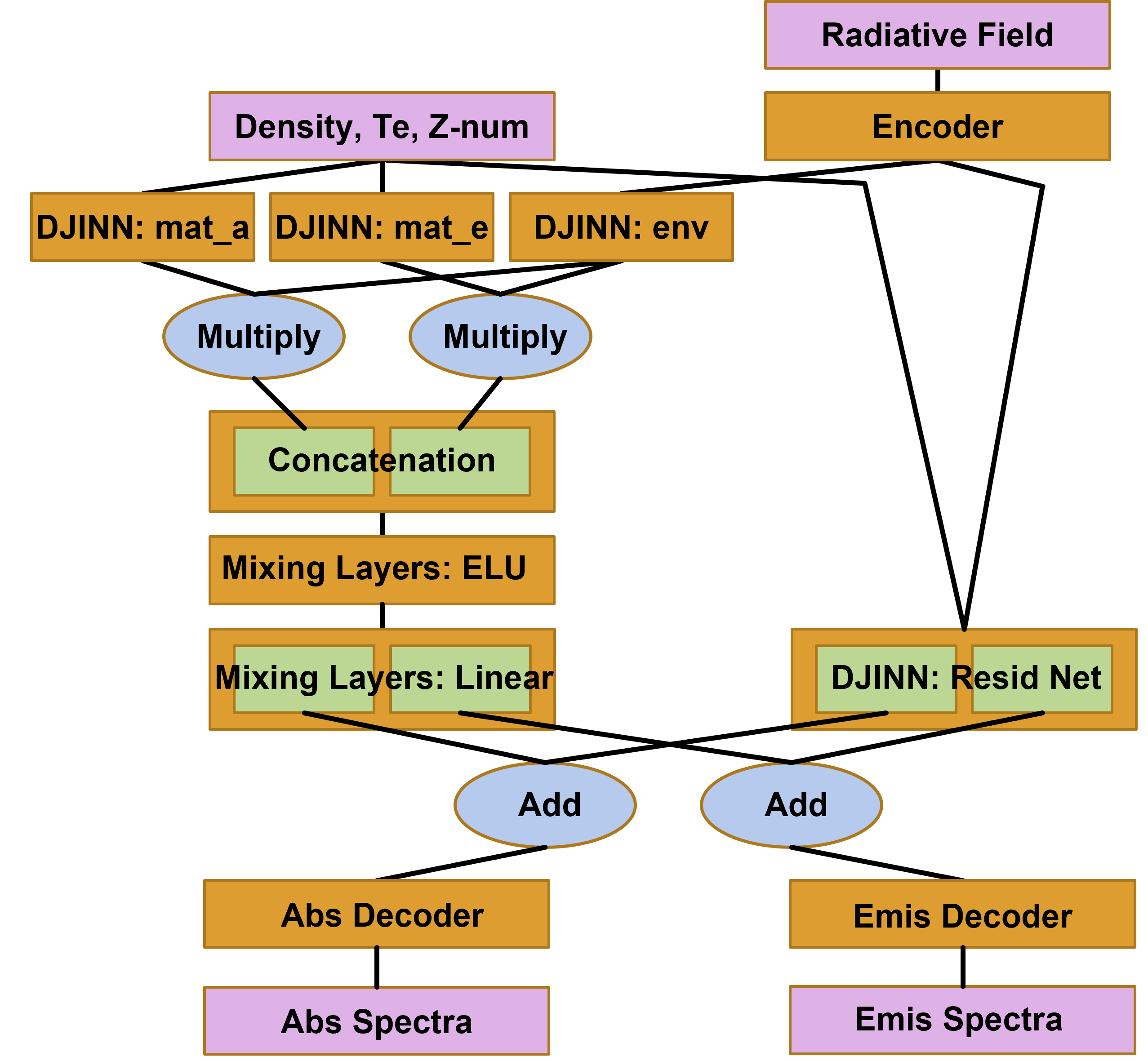}
    \caption{This is the full architecture as described. Excluded from this diagram are the locations and types of transformation operations.}
    \label{fig: full multiscale}
\end{figure}

Next, another DJINN model is made using only the radiative field as encoded by the multiscale autoencoder. It is initialized by using the encoded emissivity spectra but not actually trained intially; instead this DJINN model is placed in parallel with the other two DJINN models. The concatenated output of the three parallel models is then passed through three fully-connected layers of dimensionality equal to the concatenated output of the DJINN modules. These layers use the exponential-linear unit activation function. The output of these layers is then passed to a fully-connected layer of dimensionality equal to two times the dimensionality of the latent space of the multiscale autoencoder. This layer uses a linear activation function. All of these sub-models are combined into a single model, and the output of the model is expected to be the encoded absorptivity and encoded emissivity concatenated to each other.


The combined model is then evaluated with the training data, and the signed error is used to initialize yet another DJINN model that uses the entire input space: encoded radiative field, density, temperature, and atomic number, which are all minmax scaled, with the error in the latent space predictions being the output for initialization. No initial training is done here. The output of this DJINN model is then added to the output of the first combined model much like a residual block, and this newly combined model is then trained for 100 epochs to once again predict both the encoded absorptivity and encoded emissivity. Finally, the encoder of the multiscale autoencoder is prepended to the combined latent space prediction model, and the encoded absorptivity is passed to one instance of decoder of the multiscale autoencoder and the the encoded emissivity is passed to a duplicate instance of the same decoder such that they can be trained independently. These are all combined into one model and then trained for 10,000 epochs.

The training datasets consist of 80\% of all of the available data with the other 20\% used for validation. All training, including for the autoencoder, is done with a batch size that is 1\% of the training dataset. The learning rate for low-fidelity training is set to 0.0001. The autoencoder is trained for 2\,500 epochs. The high-fidelity transfer learning only consists of performing additional training the completed model with the high-fidelity data. The high-fidelity transfer learning is performed with a learning rate of 0.00001, and the training is performed for 1\,000 epochs. The loss function in all cases is the fraction of unexplained variance is given by 
\begin{equation}
    \mathrm{Loss} \; = \frac{\sum_{i=1}^{N} \left(y'-y\right)^2}{\sum_{i=1}^{N} \left(y'-\Bar{y}'\right)^2}.
    \label{eq: unexplained var}
\end{equation} Here $y$ is the expected value, $y'$ is the predicted value, and $\Bar{y'}$ is the mean of the predicted values. We found the use of this loss function is particularly important because it enables better shape matching of the predicted spectra.

Some additional points that must be known is that the DJINN software was modified to enable the usage of unity input and output scaling \cite{djinn_git}. The DJINN models, as previously described, use various combinations of scalings, but all outputs use unity scaling. Also, the autoencoder input and output, and thus the input and output of the full model, use 30th-root scaling as explained in (\citet{vanderwal2}).

Finally, while a single neural network can be used for each model, the authors opted to make two separate networks to constitute one neural network model. One model is for inputs with an electron temperature above 500 eV, and the other model is for inputs with an electron temperature below 500 eV. They are trained with inputs that have electron temperatures 100 eV below and 100 ev above the 500 eV split point for the ``high temperature" and the ``low temperature" models respectively. This is done so that the split point is not the edge of the model's trained input space. The 500 eV split point is chosen because there is a non-hard transition where the prediction accuracy quickly worsens with decreasing temperature that starts somewhere in the realms 300 eV to 500 eV. Emissivity prediction is particularly poor for low temperatures. Absorptivity prediction is worse at increased temperatures; however it is not necessarily the worst at the highest of temperatures. Rather, absorptivity performance is worst in a region located roughly around 900 eV to 1400 eV.

\subsubsection{Hydra}
The simulations of ICF experiments are performed using the radiation-hydrodynamics code \hydra{}. The simulations use the same setup as those performed in \citet{kluth} with the exception of the laser pulse profiles used and the binning-structure used for the spectra.

The simulations use a 1D spherical hohlraum made of krypton with 64 cells. The krypton hohlraum is set to have a density of 7.902 g/cc. Inside the hohlraum there is a fuel capsule consisting of an ablator made of polyethylene filled with deuterium-tritium ice fuel. There are various additional trace elements present in the fuel and ablator such as oxygen, germanium, and a few others. The absorptivity and emissivity of these materials were calculated using standard, tabulated values. The use of neural networks will be applied to the approximation of the krypton hohlraum.

\begin{figure}[H]
    \centering
    \includegraphics{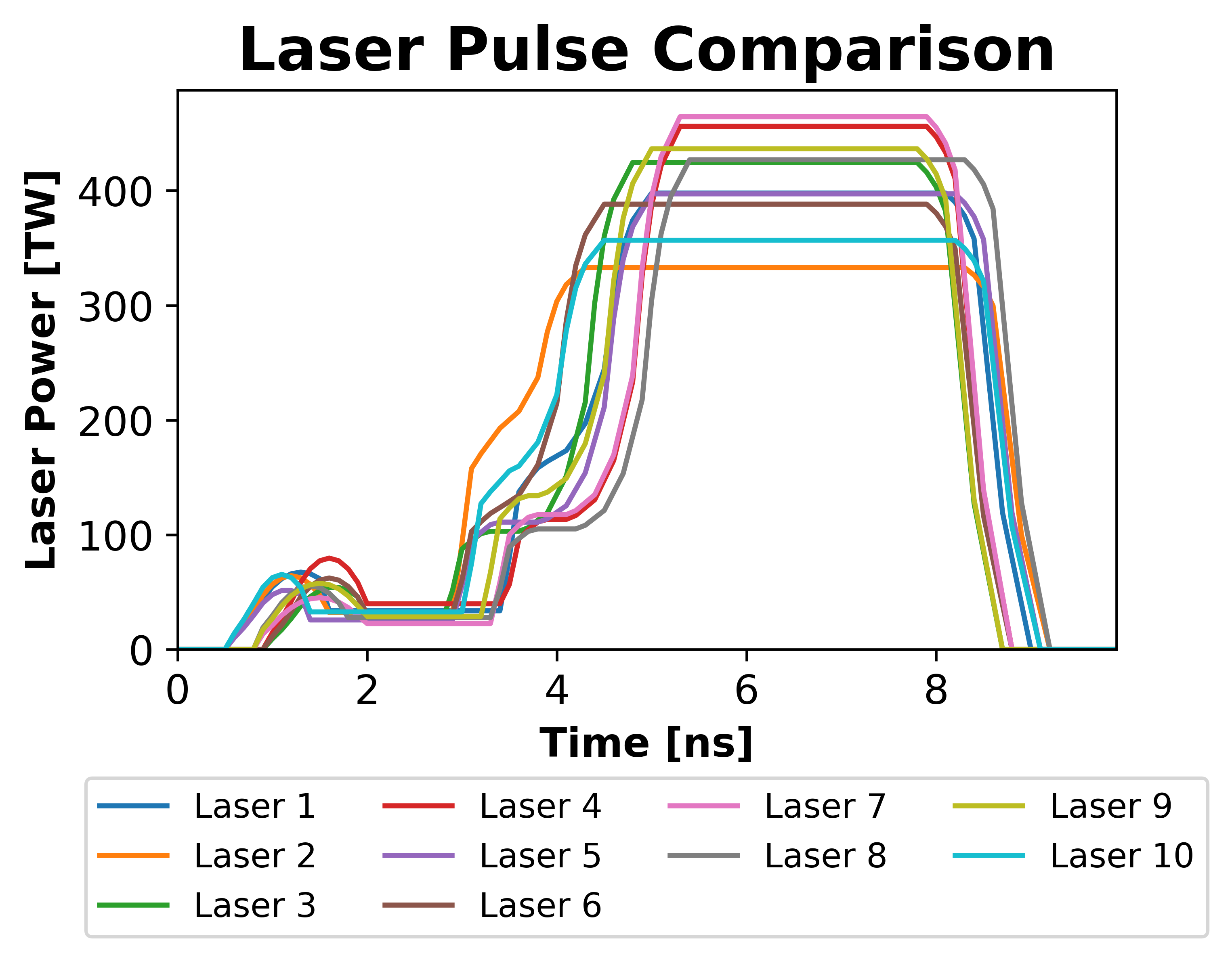}
    \caption{These are the ten laser profiles used in the simulations. They are randomly generated  perturbations to the profile of the N210808 shot at the National Ignition Facility \cite{callahan2021}}
    \label{fig: laser profiles}
\end{figure}

The laser pulse is applied as an internal source. This means that the pulse does not traverse space in the simulation and is applied to the interior cell, `cell 0,' directly. The ten laser pulse profiles that are used for both data generation and testing are shown in Figure \ref{fig: laser profiles}. During the generation of the radiative fields and a portion of the training data, the simulation of the hohlraum is handled as an NLTE problem with the radiation-hydrodynamics software \cretin{} computing the NLTE opacities and handled as a inline operation in \hydra{}. \cretin{} was set to use steady-state approximations and use Steward-Pyatt continuum lowering. When using a neural network to compute the NLTE opacities, \hydra{} passes the input data to a Python function which calculates the opacities with the neural networks.

These same ten laser profiles are used for the evaluation of the performance of the neural networks. First, all neural network models will be used in conjunction with one laser profile. Then, based on the median error in the radiative temperature, the fourth-best neural network model will be used to simulate the outcomes from all ten of the laser profiles used to produce the training data.

\section{Results}
Three simulation outputs of interest that are used to evaluate performance in the simulations are the radiative temperature, the electron or material temperature, and the density. The radiation temperature is of particular interest because it can be measured in ICF experiments, whereas the material temperature and density have to be inferred. Before getting into the simulation results, the neural networks performance is demonstrated in Table \ref{tab: errors} which shows the median values for each of the means, medians, 90th-percentiles, and maximums across all ten models. The most notable observation is that the median of the maximum $1-R^2$ value is below one for both absorptivity and emissivity for both models above and below 500 eV. This is not possible with the original method of neural network construction utilized in \cite{kluth,vanderwal1,vanderwal2}, thus the method described in section II was chosen. The median of the median $1-R^2$ value is exceptionally low as well.

\begin{table}[h]
\caption{The neural networks' performance on percent relative error and unexplained variance for absorptivity and emissivity as compared against high-fidelity data}
\label{tab: errors}
\begin{tabular}{lcccc}
\hline
\multicolumn{5}{l}{\cellcolor[HTML]{C0C0C0}\textbf{Absorptivity}}                      \\ \hline
\multicolumn{5}{l}{Binwise \% Error}                                                   \\ \hline
                                         & Mean     & Median   & 90th-Perc. & Max.     \\ \hline
\multicolumn{1}{r}{\textgreater{}500 eV} & 2.784    & 1.682    & 6.477      & 71.54    \\
\multicolumn{1}{r}{\textless 500 eV}     & 20.134   & 11.26    & 42.32      & 1151     \\ \hline
\multicolumn{5}{l}{Unexplained Variance}                                               \\ \hline
                                         & Mean     & Median   & 90th-Perc. & Max.     \\ \hline
\multicolumn{1}{r}{\textgreater{}500 eV} & 8.780e-4 & 5.342e-4 & 1.621e-3   & 7.699e-3 \\
\multicolumn{1}{r}{\textless 500 eV}     & 1.821e-2 & 5.589e-3 & 4.177e-2   & 0.3729    \\ \hline
\multicolumn{5}{l}{\cellcolor[HTML]{C0C0C0}\textbf{Emissivity}}                        \\ \hline
\multicolumn{5}{l}{Binwise \% Error}                                                   \\ \hline
                                         & Mean     & Median   & 90th-Perc. & Max.     \\ \hline
\multicolumn{1}{r}{\textgreater{}500 eV} & 7.082    & 3.826    & 14.78      & 3479     \\
\multicolumn{1}{r}{\textless 500 eV}     & 1.691e17 & 51.71    & 771.7      & 7.175e21 \\ \hline
\multicolumn{5}{l}{Unexplained Variance}                                               \\ \hline
                                         & Mean     & Median   & 90th-Perc. & Max.     \\ \hline
\multicolumn{1}{r}{\textgreater{}500 eV} & 2.603e-4 & 1.581e-4 & 4.413e-4   & 2.272e-3 \\
\multicolumn{1}{r}{\textless 500 eV}     & 1.333e-2 & 5.298e-3 & 3.126e-2   & 0.1659    \\ \hline
\end{tabular}
\end{table}

\begin{figure}[h]
    \centering
    \includegraphics[width=.48\linewidth]{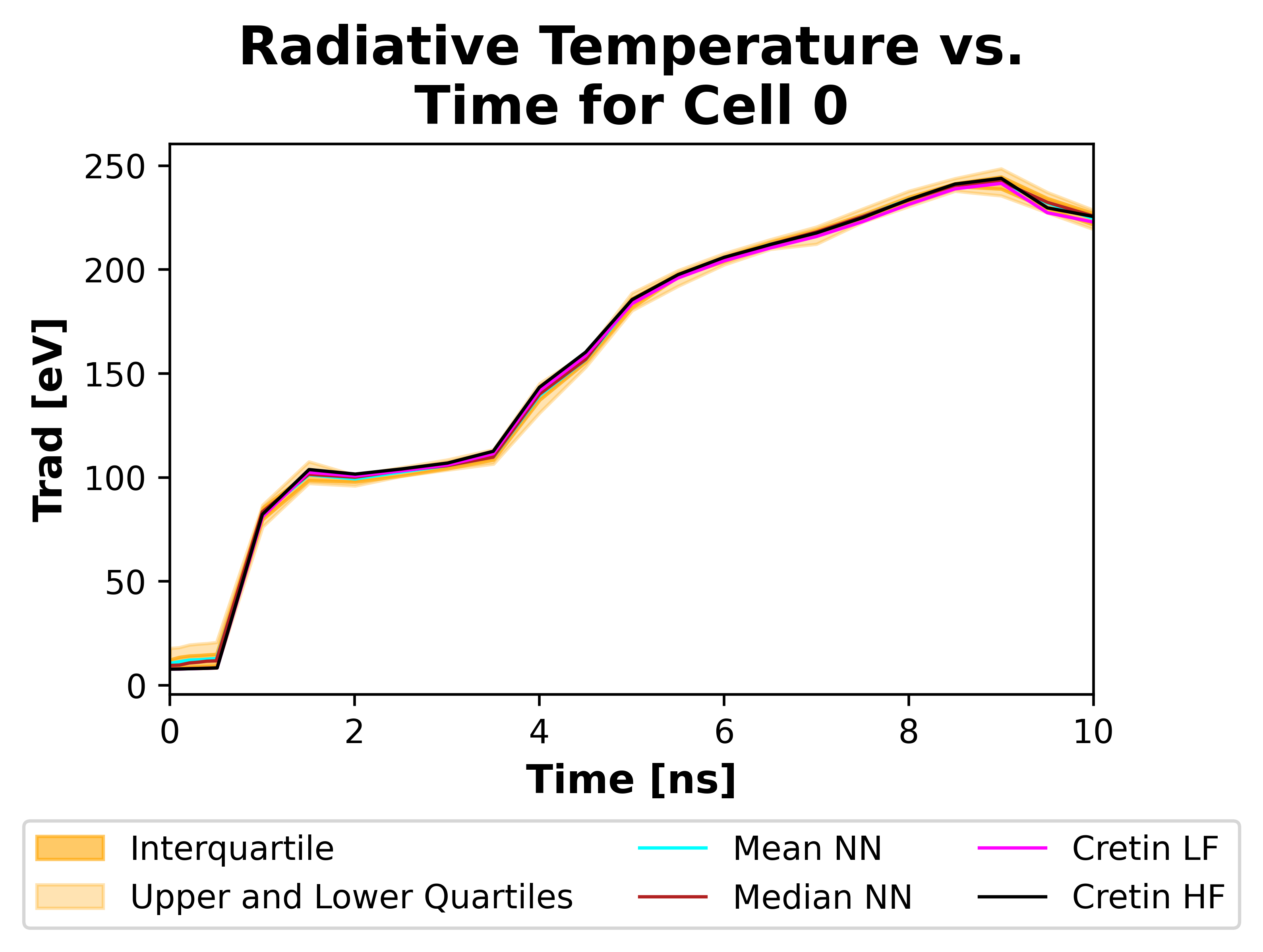}
    \includegraphics[width=.48\linewidth]{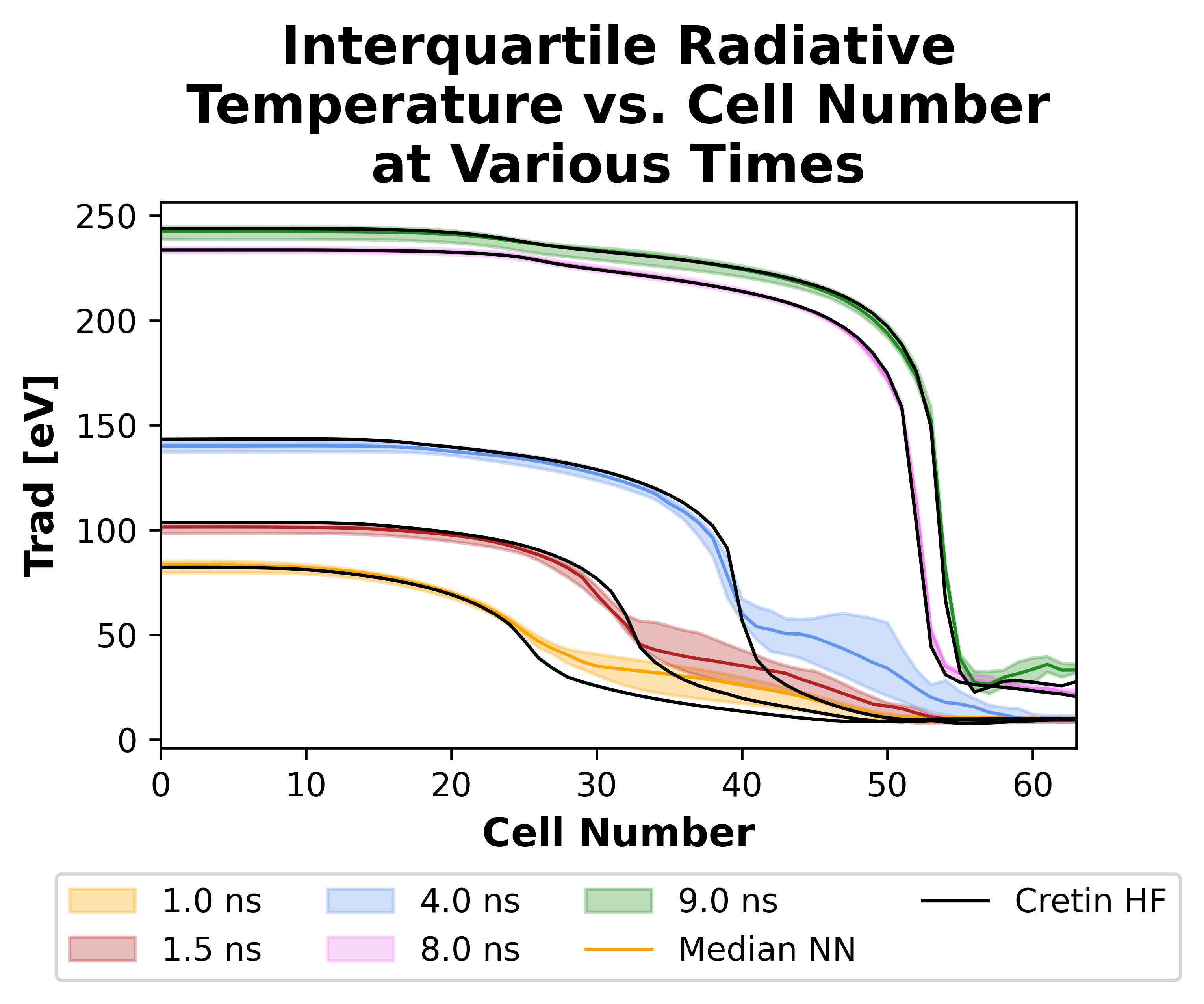}
    \caption{These plots track the radiative temperature across time for cell 0 (top) and across the cells for various times (bottom). The temperature tracking at cell 0 is quite effective; however, for other cells, the cells where the shock is present for a given time step have significant deviation from the expected values. In the bottom plot, the shaded regions represent the interquartile range with median represented by the solid line sharing the same color of the shaded region. The black lines are the expected values from the simulations.}
    \label{fig: trad interquart}
\end{figure}

In our 64 cell discretization in space, cell 0 is the surface that faces the fuel capsule and is, therefore, considered the most important cell of the simulation. The maximum radiation temperature could also be considered to be a good point of comparison for the level accuracy achieved. The left-hand plot in Figure \ref{fig: trad interquart} shows the the interquartile range as well as the ranges of the lower and upper quartiles. They are plotted against the results of  simulations ran with low-fidelity spectra and the high-fidelity spectra as well as the mean and median predicted values of the radiation temperature between the ten different neural network models. The interquartile range of the radiation temperature for each cell at various times are displayed in the right-hand plot of Figure \ref{fig: trad interquart}. Surprisingly, there is significantly higher uncertainty at earlier times and in cells between times and cells with low uncertainty. This is particularly apparent in the cells near cell 48 which have a significantly higher level of uncertainty around 4.0 ns into the simulation, yet by the end of the simulation, there is very little uncertainty in the simulation as seen in the 8.0 ns and 9.0 ns lines.

The right-hand plot also displays an apparent shift in the cell number as time advances where the median prediction begins to significantly deviate from the expected temperature profile. This deviation is expected to arise from a rapid and significant change in the radiative field intensity with the rapid change arising from the energy imparted by the laser pulse, shock heating, and the radiative field from the surrounding material. Further, the contributions from the laser and the other cells are filtered by the plasma in between the sources and a given cell. The radiative fields in this ``front'' should not prevent the implementation in future use because the rapidly transitive and evolving nature of the front makes it difficult to capture without prior knowledge, so the radiative fields in this front likely make up a small portion of the training data. Thus, the authors expect that increasing the number of samples from this phase of the simulation by either increasing the sampling rate globally for cells and time or in a more targeted fashion such as around the periods of increase in laser output power or potentially by even making a moving frame of increased sampling that follows the front.

The relative error near the peak of the radiation temperature is quite small. Interestingly, this happens to occur after having significantly higher relative error at the earlier times of the simulation. The  mean, median, and maximum percent relatives errors can be seen in \ref{tab: max trad err}. The ``time agnostic" column refers to the fact this is the error between the two maximum temperatures found anywhere in cell 0 for a given simulation. The ``time corrected" column refers to fact that this column displays the error found at the same time as the maximum radiation temperature in the high-fidelity simulation. The time-corrected median relative error comes out at 1.05\%, and the maximum time-corrected relative error comes out at 2.87\%.

\begin{table}[!t]
\centering
\caption{The percent relative error of the maximum radiation temperature in cell 0 for both time agnostic and time corrected considerations}
\begin{tabular}{rcc}
\hline
\% Error      & Time Agnostic & Time Corrected \\
\hline
mean    & 1.042          & 1.232           \\
median  & 1.049          & 1.122           \\
maximum & 1.585          & 2.871           \\
\hline
\end{tabular}
\label{tab: max trad err}
\end{table}

\begin{figure}[h]
    \centering
    \includegraphics[width=.48\linewidth]{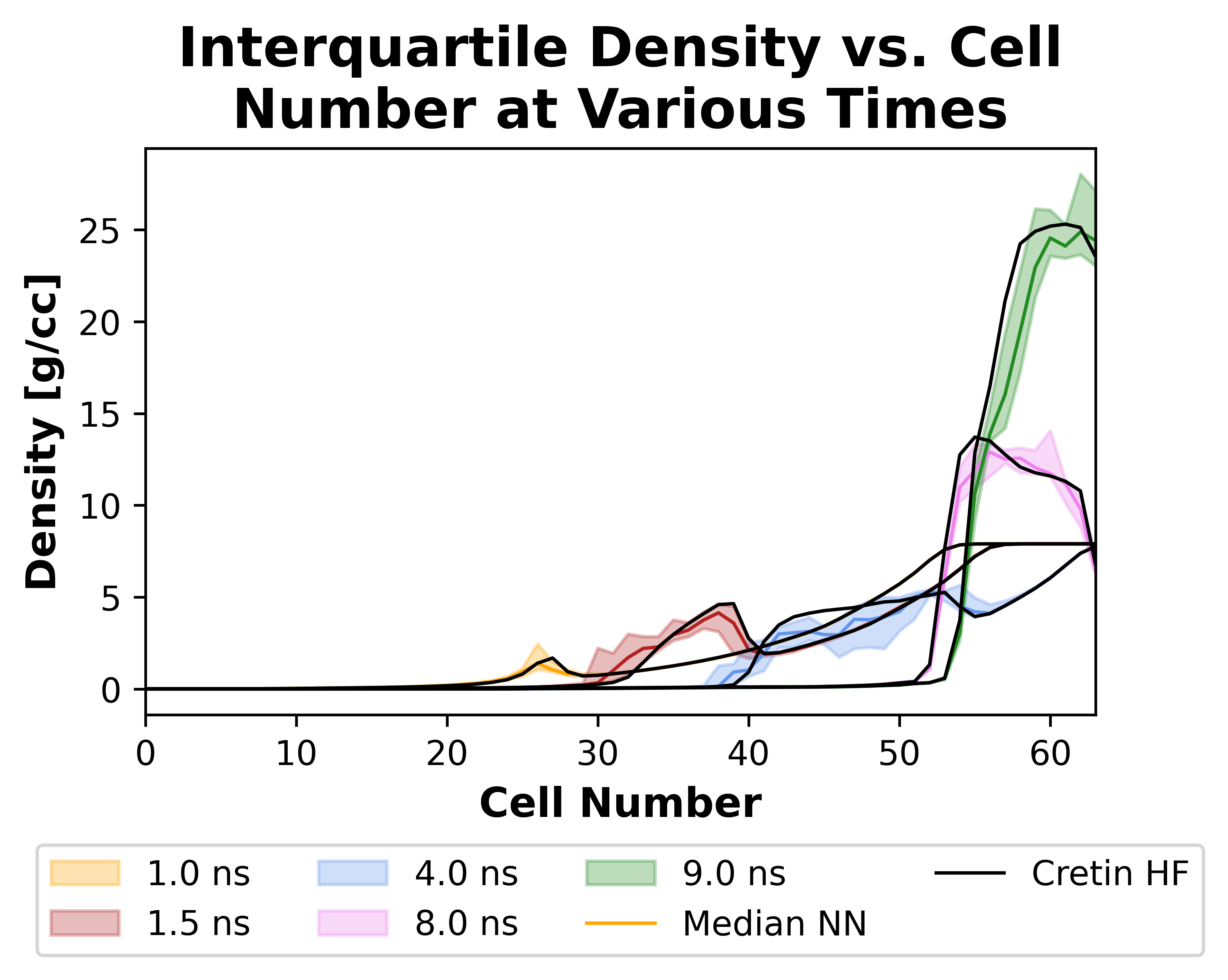}
    \includegraphics[width=.48\linewidth]{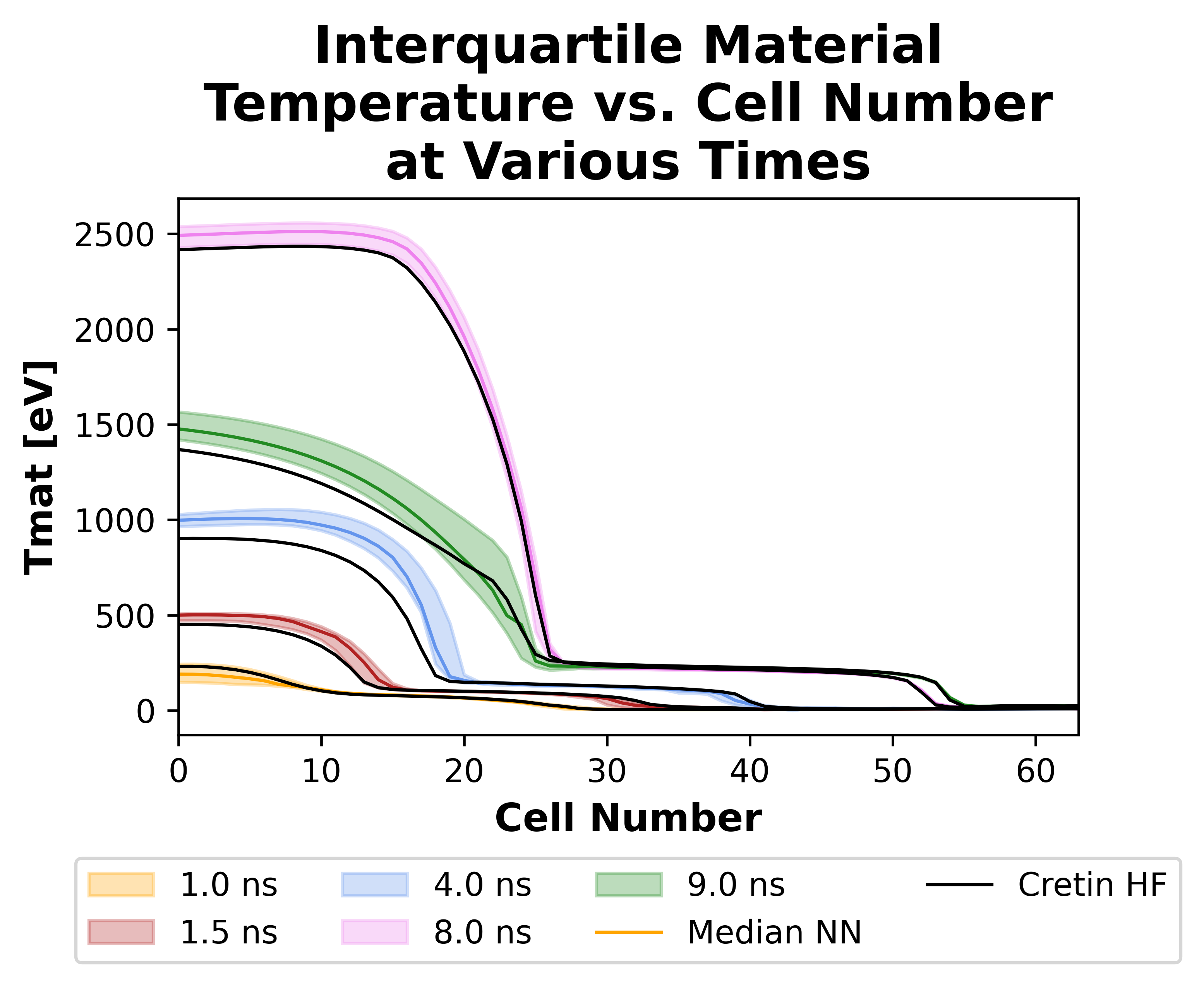}
    \caption{These plots of the density and material (electron) temperature across the cells for various times demonstrate that the density and temperature are generally predicted well when using neural networks. The density has the highest deviations near the shock, and the material temperature ahs the highest deviations near the laser source and fuel capsule. The shaded regions represent the interquartile range with median represented by the solid line sharing the same color of the shaded region. The black lines are the expected values from the simulations}
    \label{fig: den and tmat interquart}
\end{figure}

The density and material temperature  perform better in mean percent relative error than the radiation temperature. The mean error across all cells and sample times for density and material temperature are 18.7\% and 23.8\% respectively as opposed to 26.9\% for radiation temperature. However, as can be seen in the two plots in Figure \ref{fig: den and tmat interquart}, large portions of the cells for both variables see any deviation in areas that might be considered of low interest based on the baseline computation. In areas of increased activity or deviation from a seemingly constant state, the predictions become significantly worse. Additionally, just as in the radiation temperature plot in Figure \ref{fig: trad interquart}, the cell profile at 4.0 seconds shows some of the largest deviations from the expected values.

As a note of understanding what goes on during the simulation, one can see the propagation of the shock through the material in the the left-hand plot of Figure \ref{fig: den and tmat interquart} as the hump of increased density moves through the cells and actually grows over time. Similarly, in the right-hand plot, one can see how the radiation is filtered out by the plasma closest to the laser source by the fact that only the low numbered cells see significant increases in temperature. The shock heating can also be observed by comparing the location in the step in the material temperature near cell 50 at 8.0 and 9.0 seconds with the density profiles at those same times.

The second part of testing the neural networks involves the choosing the fifth and sixth best neural networks based on their mean radiation temperature error across all cells and times for laser profile 1. These two models were selected for the purpose of choosing models that may represent a more common outcome of training the neural networks. They are neither the best nor the worst models. The results are simultaneously surprising and not surprising. The mean and median error in radiation temperature across all cells, times, and laser pulses is lower for the fifth model than the sixth model with mean errors of 20.3\% and 21.4\% respectively. The median errors are even more different at 4.40\% and 7.70\%. However, when the results of Table \ref{tab: multilaser err} are considered, the time corrected error for the maximum radiation temperature of cell 0 is significantly better for the sixth model with a median error of 1.193\% compared to the 3.760\% error from the fifth model. The maximum time corrected error offers an even starker difference. That aside, if the time of the maximum temperature is not as important as simply obtaining the correct maximum temperature, the fifth model performs better with a median error of 0.3048\% compared to the 0.6414\% of the sixth model.

\begin{table}[]
\caption{The percent relative error of the maximum radiation temperature in cell 0 for both time agnostic and time corrected considerations for two models as used with simulations with each laser pulse}
\label{tab: multilaser err}
\begin{tabular}{rcccc}
\hline
\multicolumn{1}{c}{} & \multicolumn{2}{c}{Time Agnostic} & \multicolumn{2}{c}{Time Corrected} \\ \hline
\% Error                   & 5th Model       & 6th Model       & 5th Model        & 6th Model       \\ \hline
mean                 & 0.4063          & 0.5819          & 3.0945           & 1.130           \\
median               & 0.3048          & 0.6414          & 3.760            & 1.193           \\
maximum              & 1.280           & 1.043           & 5.985            & 1.729           \\ \hline
\end{tabular}
\end{table}

Lastly, the speed up in computation of opacities is considerable based on the timing of the neural network evaluation compared with inline calls from \hydra{} to \cretin{}.  For reference, the speedup over the low-fidelity simulations is roughly 2.36x or rather the neural networks consume 42.4\% of the time that the computations performed by \cretin{} do. The speed up for the high-fidelity simulation is exceptional with a speedup of 19.4x or rather neural networks consume 5.20\% of the regular computation time. An important note to make about this speedup in comparison to the speedup demonstrated by \citet{kluth} is that we are reporting the speed up in the opacity calculations only. 
The time comparisons provided in \citet{kluth} are based on running \cretin{} in a nonstandard way involving Python script calls that introduces significant overhead.  
This leads to a significant slow down in the process of using \cretin{}, and more dramatic performance gains. In our results we are reporting neural network time to standard inline \cretin{} execution. Therefore,  the speedup values should be expected to be smaller. 

\section{Conclusion}
In this work, it has been demonstrated that neural network emulators of \cretin{}  can recover the maximum radiation temperature in an integrated Hydra hohlraum simulation with a mean relative error estimated to be 1.23\%. It has also been shown that a single network from an enesmble of 10 models, can achieve a mean relative error in the maximum radiation temperature in the range 1.13\%-3.10\% across multiple laser pulses if time corrected but can achieve around 0.41\%-0.58\% if not time corrected. Moreover, this model requires only 5.20\% of the computation time compared to inline \cretin{} calculations. The density and material temperature were also demonstrated to closely match \cretin{}; however, there were regions of space and time that saw fairly significant deviation. As such, those who desire to analyze the physics further from the innermost cells will need to further improve upon these results before that can be reliably done.

An improvement in results could be achieved by using more low-fidelity data. There could also be better tuning of when the simulations are sampled for the radiative fields, absorptivity spectra, and emissivity spectra. The sampling times used here were regularly spaced, but the sampling times could  be optimized in  frequency and not necessarily be regularly spaced. The measured uncertainty with these models could also be further improved by utilizing laser profiles that were not used to generate training data unlike what was done in this work.


An alternative to a single neural network that predicts the properties of multiple chemical elements would be to use several, single element models\cite{vanderwal1}. 
This could be used to replace the fuel and fuel capsule models, which are currently based on table lookups, with neural networks might be one way multi-element models can be used. While this approach may  be slower, high-fidelity neural networks trained on more accurate NLTE data may produce more realistic simulations. Further, as expansion on previous work for multiple elements and transfer learning as well as this work, high-fidelity multi-element models could be produced that actually utilize mix ratios of multiple elements such as would be seen in the fuel and potentially fuel capsule \cite{vanderwal1,vanderwal2}.

More work could also be done on the transfer learning process which could sequentially or concurrently use multiple levels of fidelity \cite{aydin2019,meng2020}. In particular, a model that uses linear and non-linear correlations to compensate may be great choice to further expand the accuracy \cite{meng2020}. For the ambitious, physics informed neural networks (PINNs) may be an interesting path to take. This route could provide high quality results if past work is a reasonable indicator of performance; however, it would likely require a custom training loop that would have to be run in concurrently with a radiation-hydrodyanmics software package like \cretin{} \cite{meng2020,karniadakis2021,raissi2019,zhang2021,kim2022,dwivedi2020,jagtap2020,yang2019,rao2020}. Additionally, the PINNs could be used in special ways to perform transfer learning \cite{meng2020,chakraborty2021,goswami2020}.

There could also be further improvements in the architecture in ways not related to PINNs. The architecture used in this work are fully-connected with additional skip-layer features included, but a convolutional network with similar skip-layer features may perform better. There could also be an introduction more closely related to AI for the task of determining which network should take high-temperature inputs and which network should take low-temperature inputs. Rather, a network that is introduced for this reason could choose to split along a non-linear boundary that is based on all input variables. Finally, there could exist a better way to auto-construct networks than using DJINN. One such possibility is the class of evolutionary neural networks which could potentially be smaller while also better predicting than networks made with DJINN \cite{xue2021_0,xue2021_1,oneill2021,cui2019}.


\begin{acknowledgments}
This work was performed under the auspices of the U.S. Department of Energy by Lawrence Livermore National Laboratory under Contract DE-AC52-07NA27344.

This document was prepared as an account of work sponsored by an agency of the United States government. Neither the United States government nor Lawrence Livermore National Security, LLC, nor any of their employees makes any warranty, expressed or implied, or assumes any legal liability or responsibility for the accuracy, completeness, or usefulness of any information, apparatus, product, or process disclosed, or represents that its use would not infringe privately owned rights. Reference herein to any specific commercial product, process, or service by trade name, trademark, manufacturer, or otherwise does not necessarily constitute or imply its endorsement, recommendation, or favoring by the United States government or Lawrence Livermore National Security, LLC. The views and opinions of authors expressed herein do not necessarily state or reflect those of the United States government or Lawrence Livermore National Security, LLC, and shall not be used for advertising or product endorsement purposes. Released as LLNL-JRNL-834755-DRAFT.
\end{acknowledgments}

\bibliography{bib}

\end{document}